# Fully programmable slow light based on a spinor representation of generalized coupled-resonator-induced transparency

Seungkyun Park[1,2], Beomjoon Chae[2], Hyungchul Park[2], Sunkyu Yu[2§], Xianji Piao[3†], and Namkyoo Park[1*]

[1]Photonic Systems Laboratory, Department of Electrical and Computer Engineering, Seoul National University, Seoul 08826, Korea

[2]Intelligent Wave Systems Laboratory, Department of Electrical and Computer Engineering, Seoul National University, Seoul 08826, Korea

[3]Wave Engineering Laboratory, School of Electrical and Computer Engineering, University of Seoul, Seoul 02504, Korea

E-mail address for correspondence: [§]sunkyu.yu@snu.ac.kr, [†]piao@uos.ac.kr, [*]nkpark@snu.ac.kr

**Abstract**

Electromagnetically induced transparency (EIT), arising from quantum interference in coherently driven atomic systems, has inspired a variety of photonic analogues, such as coupled-resonator-induced transparency (CRIT) built on the quantum-state modelling using resonators. Although CRIT serves as a building block for slow light in photonic integrated circuits, recent advances in topological photonics motivate a further generalization of both EIT and CRIT using gauge-field degrees of freedom. Here, we propose generalized CRIT via a spinor representation with dual-channel gauge fields, enabling fully programmable CRIT featuring dynamical spectral engineering. We generalize the traditional EIT framework by introducing a spinor representation of bright- and dark-mode resonances, yielding a unified description of design parameters through universal unitary operations. Implementing a coupled-resonator building block that accesses the entire design space through dual-channel gauge fields, we demonstrate a programmable slow-light band in a one-dimensional CRIT lattice. These results address urgent needs in optical interconnects, such as tunable delay lines, reconfigurable synchronization, and linear frequency conversion.

## Introduction

Electromagnetically induced transparency (EIT) is a representative example of quantum interference between states in atoms or molecules, typically occurring in three-level systems[1,2] referred to as the Λ-type configuration[3]. The major impact of EIT lies in dramatically modified optical nonlinearities in coherently prepared media[4,5], which have been employed in activation functions for optical neural networks[6], CNOT gates for quantum computing[7], and quantum memories[8]. Although EIT belongs to atomic, molecular, and optical (AMO) physics, its classical counterparts have profoundly enriched the landscape of coherent wave manipulation. For example, emulating the Λ-type quantum state configuration across coupled oscillators[9], plasmonic structures[10,11], acoustoelastic cavities[12], and metamaterials[13,14], has unlocked sophisticated spectral engineering by mimicking EIT responses, such as slow- and stopping-waves[12,15,16]. Especially, the framework of coupled-resonator-induced transparency (CRIT)[16-19] establishes systematic spectral manipulations within photonic integrated circuits.

The traditional Λ-type configuration for EIT has been generalized into extended schemes in AMO physics[3] to achieve enhanced nonlinearities, such as giant Kerr effects[20] and strong four-wave mixing[21]. In the context of classical quantum simulations[22], recent progress in integrated photonics, which has harnessed extended wave degrees of freedom—including matrix-valued gauge fields[23], synthetic dimensional lattices[24], non-Hermitian systems[25], and system reconfigurability[26]—motivates the further generalization of EIT for photonic circuit functionalities. Specifically, this approach will impose novel design freedom on slow light and its applications to tunable optical buffers and photonic memories essential for optical interconnects and quantum computing[27-30].

Here, we propose a programmable CRIT platform that transcends the conventional Λ-type configuration. We generalize the theoretical framework for the EIT and CRIT through a spinor representation of bright- and dark-mode resonances, enabling their coupling to be interpreted via SU(2) operations in conjunction with bright-mode decay. To realize the proposed framework within an integrated photonic platform, we devise a fundamental building block comprising resonators coupled to a waveguide, where the resonators are coupled via reconfigurable loop couplers—the design that facilitates comprehensive control over inter-resonator coupling. Leveraging this architecture and its lattice extension, we demonstrate the on-demand tailoring of EIT-analogous spectral features—including linewidth, spectral asymmetry, and lattice dispersion—which can be decoupled through individual SU(2) operations. Alongside the demonstrated reconfigurable slow light with linear frequency conversion, our results significantly expand the design degrees of freedom for programmable photonic circuits, enabling dynamical functionalities, such as synchronization and frequency conversion.

## Results

### CRIT generalization

We revisit the CRIT—an optical analogy of EIT phenomena in integrated photonics[11,15-19,31-34]. The conventional EIT system is realized by the interactions between two quantum states—the bright ($|\psi_B\rangle$) and dark ($|\psi_D\rangle$) states—which have shorter and longer lifetimes, respectively[1,2]. The system induces the destructive quantum interference between the probe ($|0\rangle\rightarrow|\psi_B\rangle$) and control ($|\psi_D\rangle\rightarrow|\psi_B\rangle$) transitions, which suppresses absorption through the probe transition (Fig. 1a), and therefore, leads to the transparency within the absorption spectrum. In the corresponding CRIT, the lifetime discrepancy between $|\psi_B\rangle$ and $|\psi_D\rangle$ is modelled by different quality ($Q$-) factors of optical resonators. The interactions between $|\psi_B\rangle$ and $|\psi_D\rangle$ resonators, which have resonance frequencies of $\omega_0 + \Delta\omega$ and $\omega_0 - \Delta\omega$, respectively (Fig. 1b), are characterized by the coupling coefficient $\kappa_{BD}$ between the resonators. Through two distinct coupling paths to the bright-mode resonator from the dark-mode resonator and from a probe waveguide, the transmission along the waveguide is controlled by the wave interference at the bright-mode resonator. Notably, the emergence of the CRIT is robust against the resonant frequency detuning $2\Delta\omega$, while imposing Fano spectral asymmetry on the CRIT spectrum[11].

In interpreting this conventional picture of CRIT, we employ the spinor representation to generalize the interaction between bright and dark modes. We define a two-level spinor state $|\Psi\rangle = [\psi_B, \psi_D]^T$, where $\psi_B$ and $\psi_D$ are the complex-valued amplitudes of $|\psi_B\rangle$ and $|\psi_D\rangle$ resonator modes, respectively. By assigning intrinsic lifetimes of $\tau_B^I$ and ideally infinity to the bright and dark modes, respectively, we formulate the governing equation for the system in Fig. 1a,b—excluding the excitation waveguide—as follows:

$$i\frac{d}{dt}|\Psi\rangle = -\left[\left(\omega_0 + \frac{i}{4\tau_B^I}\right)I + \kappa_{BD}\sigma_x + \left(\Delta\omega + \frac{i}{4\tau_B^I}\right)\sigma_z\right]|\Psi\rangle \triangleq \left(H_0^H + H_0^A\right)|\Psi\rangle, \qquad (1)$$

where $\sigma_{x,y,z}$ denote the Pauli matrices, $I$ is the identity matrix, and $H_0^H = -(\omega_0 I + \kappa_{BD}\sigma_x + \Delta\omega\sigma_z) = (H_0^H)^\dagger$ and $H_0^A = -i(I + \sigma_z)/(4\tau_B^I) = -(H_0^A)^\dagger$ are the Hermitian and anti-Hermitian parts of the Hamiltonian, respectively. Equation (1) shows that the CRIT Hamiltonian leads to the $H_0^H$-induced rotation operations around the $x$- and $z$-axes on the Bloch sphere of $|\Psi\rangle$, while possessing the $H_0^A$-induced anti-Hermiticity of the same coefficient $1/(4\tau_B^I)$ on the $z$-axis rotation and the global phase evolution. While the $z$-axis anti-Hermiticity imposes differentiated lifetimes on the bright- and dark-mode resonators, distinct rotations about $x$- and $z$-axes control the interference and the corresponding CRIT.

At first glance, the operations dictated by $H_0^{\mathrm{H}}$ might appear to satisfy the requirements for universal SU(2) operations, according to established criteria for universal gates[26,35]. However, because evanescent coupling $\kappa_{\mathrm{BD}}$ is primarily governed by the physical proximity of the resonators—which remains invariant under a stationary geometry—the dynamics accessible via $H_0^{\mathrm{H}}$ are confined to a specific subset of the SU(2) group due to the almost fixed $x$-axis rotation. Therefore, the proposed spinor representation raises a conceptual challenge: the generalization of CRIT—and more generally EIT—into the form of a universal SU(2) operation combined with the decay contrast from the anti-Hermitian part $H_0^{\mathrm{A}}$.

To address this challenge, we propose the generalized EIT picture (Fig. 1c) and its corresponding CRIT platform (Fig. 1d), achieving programmable and universal SU(2) operations under the stationary distributions of photonic elements. To achieve the universal SU(2) operations and the reconfigurability at the same time, we employ the off-resonant loop couplers $|\mu^{\mathrm{U}}\rangle$ and $|\mu^{\mathrm{L}}\rangle$ (Fig. 1c,d), which provide the indirect coupling with dual-channel gauge fields[36]. The proposed platform, again except for the excitation waveguide for simplicity, is governed by the equation, $i(d/dt)|\Psi\rangle = (H^{\mathrm{H}} + H_0^{\mathrm{A}})|\Psi\rangle$, where the Hermitian part $H^{\mathrm{H}}$ becomes (Supplementary Note S1):

$$H^{\mathrm{H}} = -\left(\omega_0 I + \frac{1}{2}\sqrt{\frac{1}{\tau_{\mathrm{D}}\tau_{\mathrm{B}}}}\left[\left(\cos\xi^{\mathrm{U}} + \cos\xi^{\mathrm{L}}\right)\sigma_x + \left(\sin\xi^{\mathrm{U}} + \sin\xi^{\mathrm{L}}\right)\sigma_y\right] + \Delta\omega\sigma_z\right), \quad (2)$$

while $\xi^{\mathrm{U,L}}$ are the phase shifts through each loop coupler, and $\tau_{\mathrm{B,D}}$ are the lifetimes of the bright- and dark-mode resonators with respect to the couplers, respectively. We note that while the $z$-axis rotation can be tailored with the tuning of resonant frequencies, two degrees of freedom $\xi^{\mathrm{U,L}}$ that originate from the tunable phase modulations allow for the independent control of $x$- and $y$-axis rotations. Therefore, the platform described by Eq. (2) generalizes the CRIT phenomena to the programmable and universal SU(2) operations, maintaining the uniqueness of the EIT: anti-Hermiticity on the $z$-axis rotation and the global phase evolution.

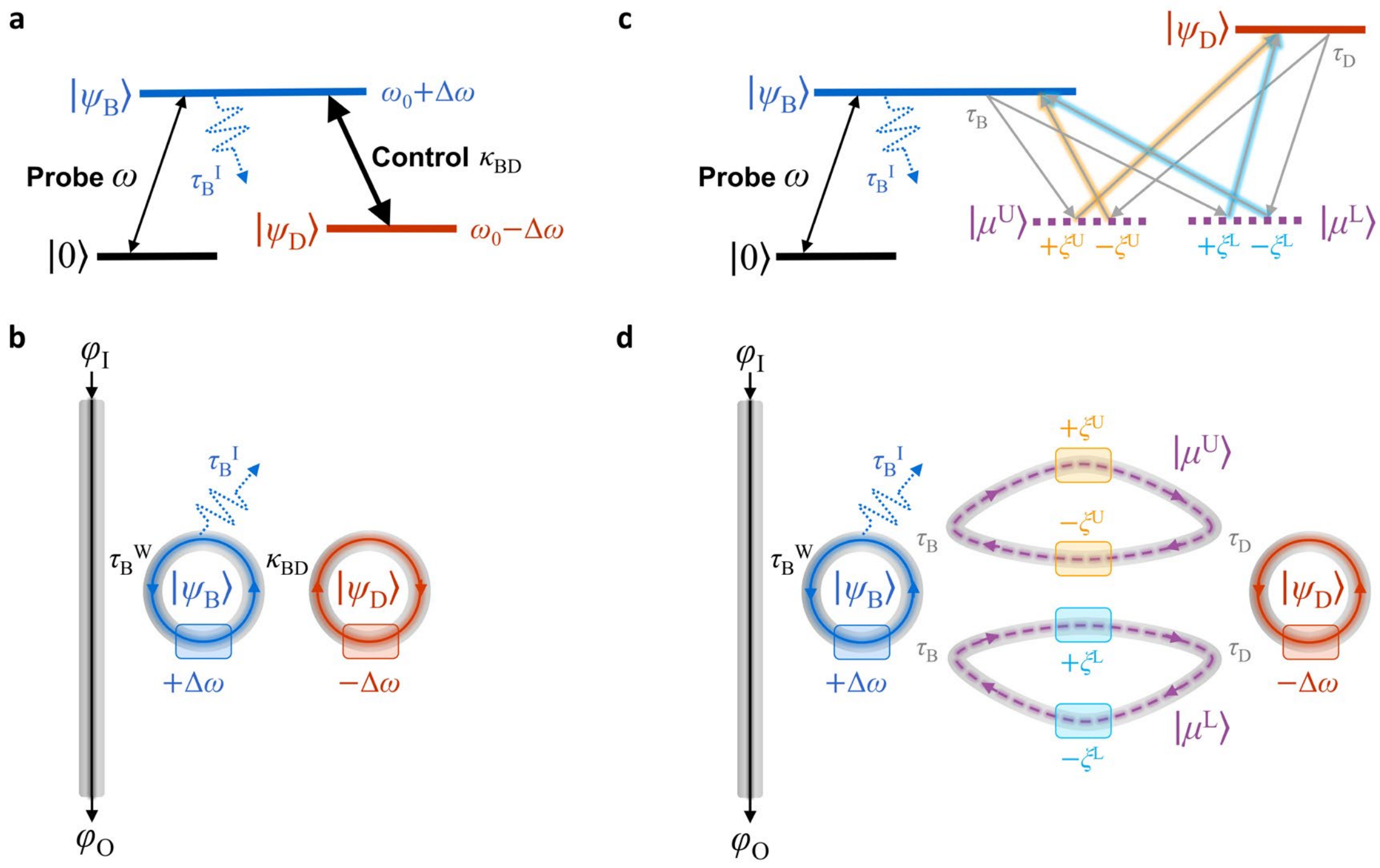

**Fig. 1. Spinor generalization of EIT-CRIT systems. a,b,** Traditional and **c,d,** generalized configurations: **a,c,** Energy-level diagrams and **b,d,** the corresponding CRIT platform schematics. The blue and red colours represent the bright ($|\psi_B\rangle$) and dark modes ($|\psi_D\rangle$), respectively, with the resonance frequencies $\omega_0 \pm \Delta\omega$. In (**c**,**d**), a pair of off-resonant loop coupler states ($|\mu^U\rangle$, $|\mu^L\rangle$) is employed (dashed purple arrows), where $\xi^U$ and $\xi^L$ are acquired through the coupling. The input and output waveguide fields are denoted as $\varphi_I$ and $\varphi_O$, respectively in (**b**,**d**). $\tau_B{}^W$ denotes the coupling to the external waveguide.

## Engineering generalized CRIT

To investigate the accessible functionalities of the generalized CRIT, we examine its SU(2) operations in terms of the $x$-, $y$-, and $z$-axis rotation operations on the Bloch sphere (Fig. 2a-c) manipulated by $\xi^{U,L}$ and $\Delta\omega$. Because only the $z$-axis exhibits anti-Hermiticity, we distinguish the effects of rotations about the $x$- and $y$-axes from those about the $z$-axis. First, the $x$- and $y$-axis rotations on the Bloch sphere (Fig. 2a,b) are governed by the Hamiltonian $H^{\mathrm{H}}$ in Eq. (2) with the conditions of $\xi^{\mathrm{U}} = -\xi^{\mathrm{L}} \equiv \xi$ and $\xi^{\mathrm{U}} = \pi - \xi^{\mathrm{L}} \equiv \xi$, respectively, under $\Delta\omega = 0$:

$$H^{\mathrm{H}} = -\left[\omega_0 I + \left(\sqrt{\frac{1}{\tau_{\mathrm{D}}\tau_{\mathrm{B}}}}\cos\xi\right)\sigma_x\right] \quad (x\text{-axis rotation}),$$
$$H^{\mathrm{H}} = -\left[\omega_0 I + \left(\sqrt{\frac{1}{\tau_{\mathrm{D}}\tau_{\mathrm{B}}}}\sin\xi\right)\sigma_y\right] \quad (y\text{-axis rotation}). \qquad (3)$$

The trigonometric coefficients of $\sigma_x$ and $\sigma_y$ determine the rates of the coherent mixing between the bright and dark states due to the off-diagonal forms of $\sigma_{x,y}$. Because these transition rates determine the strength of the wave interference, $\xi$ operates as the design parameter for the CRIT spectral bandwidth (Fig. 2a,b; Methods). To quantify the bandwidth engineering through $\xi^{\mathrm{U,L}}$, we analytically calculate the full width at half maximum (FWHM) $\Gamma$ of the CRIT spectrum at $\Delta\omega = 0$ (Fig. 2d; Methods).

In contrast to rotations around the $x$- and $y$-axes, spinor evolution around the $z$-axis aligns with anti-Hermiticity. While maintaining the differentiated lifetimes of the bright- and dark-mode resonators, tailoring $\Delta\omega$ for the $z$-axis rotation affects the interference between these modes established by the $x$- and $y$-axis rotation behaviours. This altered interference is represented by Fano spectral asymmetry[11] along the frequency axis (Fig. 2c), which can be quantified by the spectral asymmetry factor $F$ (Fig. 2e; Methods). We emphasize that by tailoring $\xi^{\mathrm{U,L}}$ and $\Delta\omega$, which can be controlled in integrated photonic platforms, the observed engineering of the transmission band through universal SU(2) operations can be achieved in a reconfigurable manner. This universal nature provides the necessary degrees of freedom for the decoupled tailoring of spectral bandwidth and asymmetry, enabling versatile spectral reshaping that remains unattainable within conventional CRIT or existing EIT-like photonic platforms.

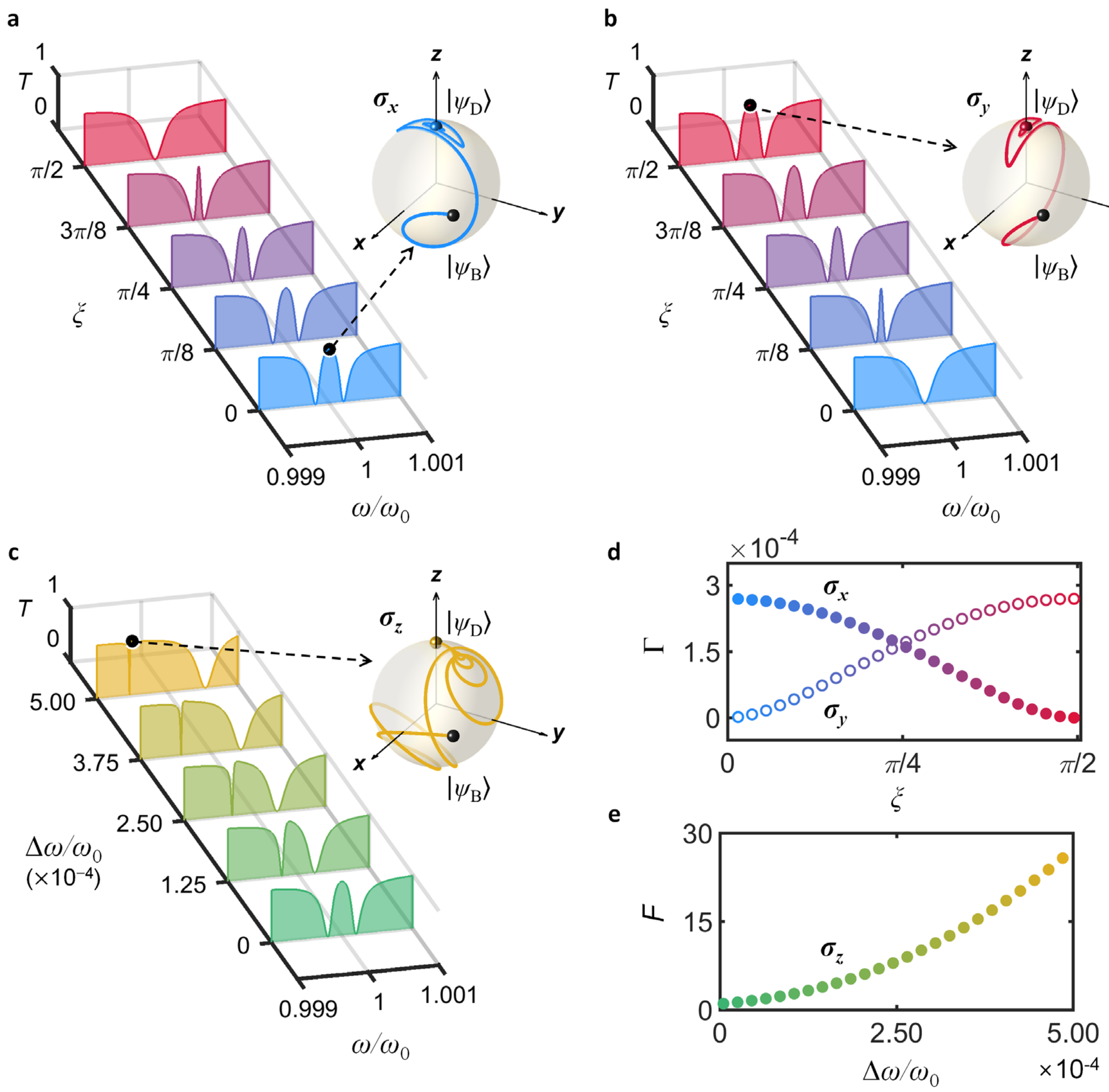

**Fig. 2. Engineering the CRIT spectra. a-c,** Transmission ($T$) spectra and state trajectories under (**a**) $x$-, (**b**) $y$-, and (**c**) $z$-axis rotations of $|\Psi\rangle$. Each plot shows $T$ as a function of $\omega/\omega_0$, with sweeps over $\xi \in [0, \pi/2]$ in **a**,**b**, and over $\Delta\omega/\omega_0 \in [0, 5.00 \times 10^{-4}]$ in **c**. The Bloch sphere illustrates the spinor evolution from the initial state (black dot, $|\Psi_0\rangle = [(1+i)/2, 1/\sqrt{2}]^{\mathrm{T}}$) to final state (coloured dots) at a given CRIT state (dashed arrows). The trajectories represent the evolution over $t \in [0, 20{,}000 \times 2\pi/\omega_0]$. **d**, FWHM as a function of $\xi$. **e**, Spectral asymmetry factor $F$ as a function of $\Delta\omega$ for $z$-axis rotations. $F = 1$ corresponds to a symmetric spectrum, while deviations from unity indicate increasing spectral asymmetry. The simulations were conducted using $\tau_{\mathrm{B}}$ and $\tau_{\mathrm{D}}$ set to 800 $\times$ $2\pi/\omega_0$, and $\tau_{\mathrm{B}}^{\mathrm{I}}$ and $\tau_{\mathrm{B}}^{\mathrm{W}}$ set to 2,000 $\times$ $2\pi/\omega_0$. In (**c**), $\xi^{\mathrm{U}} = \xi^{\mathrm{L}} = -\pi/8$.

**Programmable CRIT lattices**

Utilizing the CRIT building blocks, we implement one-dimensional (1D) lattices to investigate their slow-light bands. Figure 3a shows a waveguide coupled to a periodic array of the generalized CRIT building blocks with a periodicity of $\Lambda$. The dispersion relations and the consequent group velocities for various sets of system parameters are calculated using the temporal coupled mode theory[37,38] in conjunction with the Bloch theorem and the lossless condition of $\tau_B^I \rightarrow \infty$ (Supplementary Note S2). To explore the spectral engineering of the bands, we examine two distinct operation modes: $x$-axis rotations by tuning $\xi$ while maintaining $\Delta\omega = 0$ (Fig. 3b,c) and $z$-axis rotations with controlling $\Delta\omega$ for $\xi = \pi/4$ (Fig. 3d,e). In both cases, the CRIT lattices exhibit gapless bands, because the resonators side-coupled to the waveguide operate as phase shifters with unity transparency[39,40].

In the operation mode with $x$-axis rotations, the linewidth control for the building block derives the corresponding band engineering of the CRIT coupled-resonator optical waveguide (CROW) band[41], as shown in Fig. 3b. We note that light slows down at the dips of the CRIT spectra shown in Fig. 2a; two slow-light bands appear at $\xi = 0$ and $\xi = \pi/4$, separated by the CRIT peak, however, only a single, less pronounced slow-light band remains at $\xi = \pi/2$ upon the annihilation of the CRIT peak. Inheriting the characteristics of the CROW band[41], the group velocity of the CRIT band (Fig. 3c) decreases near the Brillouin-zone edge despite the increasing group velocity dispersion (GVD).

In addition to the bandwidth engineering, the operation mode with $z$-axis rotations provides a further degree of freedom for dispersion band shaping. As shown in Fig. 3d, the tunable Fano spectral asymmetry observed in Fig. 2c leads to the reconfigurable asymmetry in the dispersion band and the consequent group velocity (Fig. 3e). Furthermore, owing to the non-Lorentzian line shape of the Fano-asymmetric EIT band, the second operation mode enables the GVD manipulation at the target frequency. Therefore, the proposed lattice structures allow for the simultaneous engineering of the first- ($\partial k/\partial \omega$) and second-order ($\partial^2 k/\partial \omega^2$) derivatives of the dispersion through the programmable system parameters, which is an essential feature of buffers and delay lines[42,43] for optical interconnects (see Supplementary Note S3 for dispersion relations incorporating material loss).

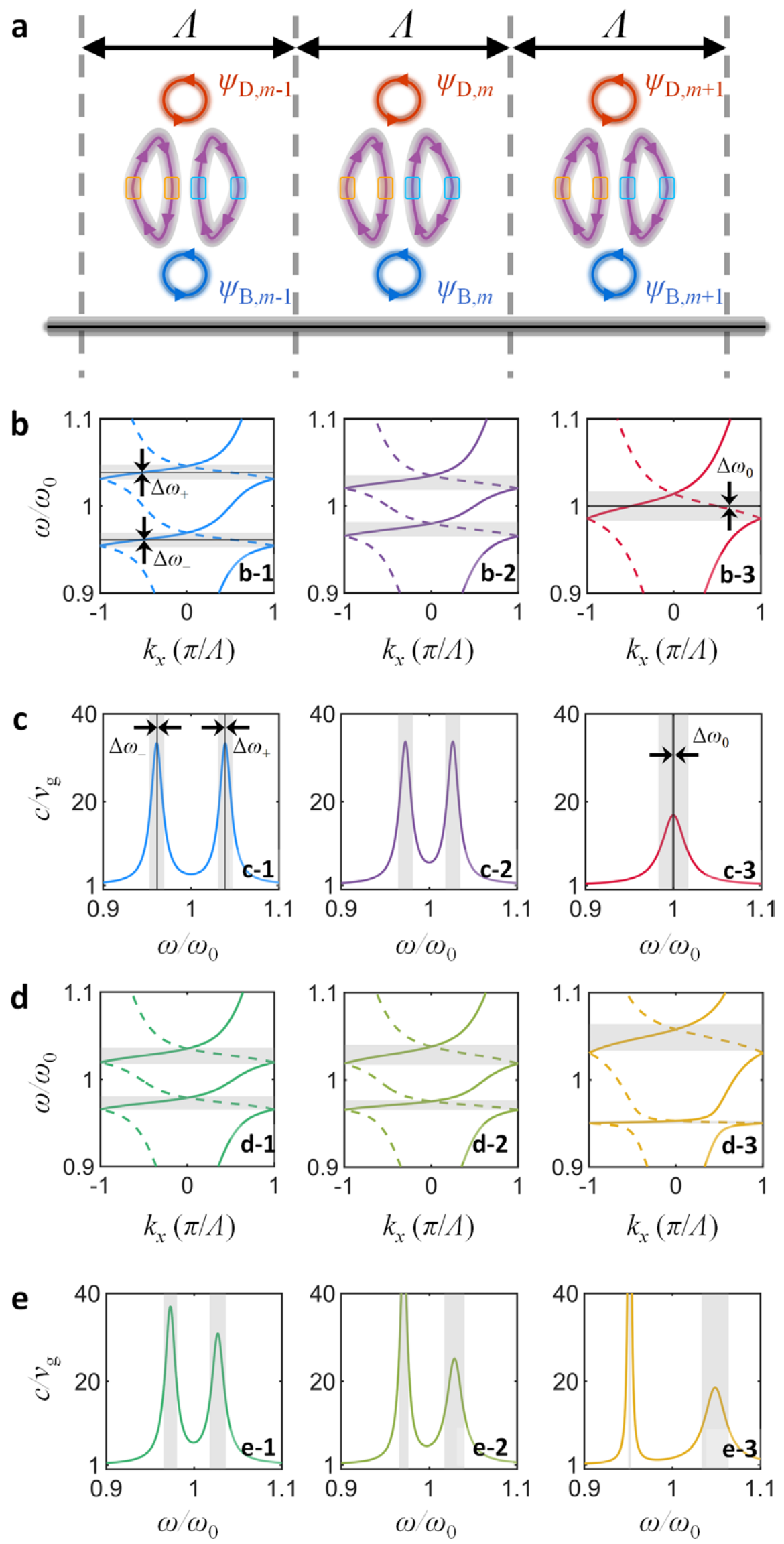

**Fig. 3. Programmable CRIT lattices. a,** Schematic of the lattice. **b-e,** Dispersion relations (**b,d**) and the group indices (**c,e**) under *x*-axis (**b,c**) and z-axis (**d,e**) rotations. The columns in **b**,**c** correspond to $\xi$ values of 0 (left), $\pi/4$ (centre), and $\pi/2$ (right) with $\Delta\omega = 0$. The columns in **d**,**e** correspond to $\Delta\omega/\omega_0$ values of $2.50 \times 10^{-3}$ (left), $1.00 \times 10^{-2}$ (centre), and $4.00 \times 10^{-2}$ (right) with $\xi = \pi/4$. In **b**,**d**, solid and dashed curves indicate forward and backward propagating modes, respectively. The shaded regions indicate the slow-light bandwidths. Arrows in **b**,**c**, depicting $\Delta\omega_0/\omega_0 = 2.65\times10^{-3}$, and $\Delta\omega_\pm/\omega_0 = 1.36\times10^{-3}$, present the input and output bandwidths employed in Fig. 4. $\Lambda = 5\pi c/(2\omega_0)$, $\tau_B = \tau_D = 4 \times 2\pi/\omega_0$, and $\tau_B{}^W = 10 \times 2\pi/\omega_0$.

**Reconfigurable slow light dynamics**

Based on the dispersion engineering shown in Fig. 3, we demonstrate the reconfigurable slow-light operation in a CRIT lattice. We investigate a 60-cell lattice with a total length of $L = 59\Lambda$, where the position between the $p$-th and the ($p$+1)-th cells is defined as $L_p = p\Lambda$. The slow-light dynamics are analysed by launching an $\omega_0$-centred Gaussian pulse into the waveguide, with its spectral width matched to the CRIT bandwidth at $\xi = \pi/2$ (right panels of Figs. 3b and 3c). The resulting dynamics are obtained by solving the temporal coupled mode equations using the 4th-order Runge–Kutta scheme (see Methods).

In evaluating the pulse evolution under temporal control of $\xi$ across the lattice, we focus on the instantaneous modulation from $\xi = \pi/2$ to $\xi = 0$ for the $x$-axis rotation ($\Delta\omega = 0$). This operation corresponds to the transition from weak to strong bright-dark-mode couplings in the CRIT configuration (Fig. 2a), achieving the switching from one broadband state to two narrow-band slower states (left panels in Figs. 3b and 3c).

Figure 4a presents the numerically calculated group index, $n_g = c/v_g$, derived from the temporal dynamics observed in Fig. 4b. The local group velocity is calculated by extracting the temporal centre of mass (CoM) at each unit cell—defined as $t_p = \langle t(L_p)\rangle = \int t|\varphi_O(t; L_p)|^2 dt / \int|\varphi_O(t; L_p)|^2 dt$ for the $p$-th cell—where $\varphi_O(t; L_p)$ represents the time-varying output field at $L_p$. Based on the temporal CoM, the local group velocity is evaluated as $v_g(L_p) = \Lambda/(t_{p+1} - t_p)$. The result in Fig. 4a indicates that the group index increases from approximately 16 to 33—corresponding to a twofold deceleration of the pulse—which is in excellent agreement with the dispersion band of Fig. 3c. We note that the response time for the instantaneous modulation (indicated by dashed lines in Fig. 4a) reflects the characteristic time required for each unit cell to establish CRIT dynamics, which leads to the bandwidth-dependent nature of the modulation speed (see Supplementary Note S4).

We note that the modulation across the entire system results in the breaking of time-translational symmetry[44-46], which relaxes the requirement for energy conservation while preserving the momentum component of the pulse during frequency conversion[47]. This operation principle is demonstrated by spectral evolutions across the lattice (Fig. 4c), showing the successful time-frequency reshaping from an $\omega_0$-centred band into two sidebands at $1.04\omega_0$ and $0.96\omega_0$. While this linear frequency conversion highlights the potential of the CRIT platform for broadband frequency converters, extending the modulation from Hermitian to non-Hermitian regimes could enable the implementation of nonreciprocal broadband frequency conversion[48] in linear systems.

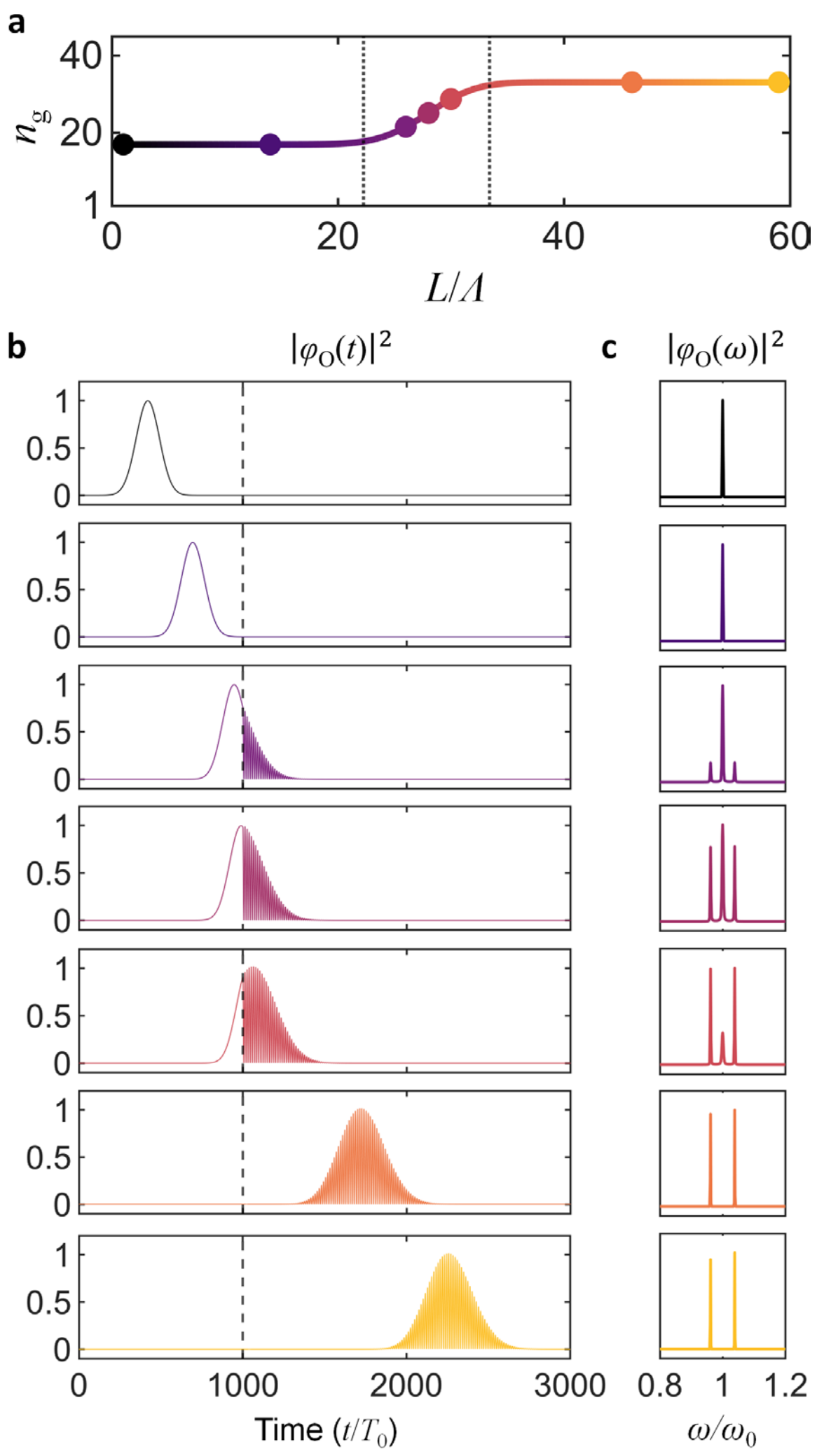

**Fig. 4. Dynamical engineering of slow light using programmable CRIT. a,** Group index $n_g$ evaluated at each unit cell along the 1D programmable CRIT lattice. The gauge field $\xi$ is switched from $\pi/2$ to 0 at $t = 1000T_0$. The coloured markers indicate the positions ($p$ = 1, 14, 26, 28, 30, 46 and 59) plotted in **b** and **c**. **b,** Temporal evolution of the pulse $|\varphi_O(t)|^2$ at the selected positions in **a**. The dashed vertical line indicates the modulation. **c,** Frequency spectra $|\varphi_O(\omega)|^2$ for the temporal traces in **b**. All the other parameters are the same as those in Fig. 3.

## Discussion

Because we examine slow-light phenomena assuming a relatively large magnitude of coupling between bright- and dark-mode resonators considering the cost for analysing temporal dynamics, further analysis is necessary regarding practical implementations of our CRIT platforms. When harnessing conventional programmable photonic circuits under linear optical modulations, we can envisage the use of silicon nitride ($Si_3N_4$)[49] or lithium niobate on insulator (LNOI)[50] platforms, allowing for intrinsic $Q$-factor resonators exceeding $10^6$—approximately, $\tau > 10^6/\omega_0$—in the footprint below 0.25 mm$^2$ at the telecom wavelength. In such high-$Q$ platforms, one can employ electro-optic[51-53], micro-electromechanical systems[54] (MEMS), or thermo-optic[55] tuning as linear optical modulations, which enables the resonance detuning of the ratio from $10^{-4}$ to $10^{-3}$, guaranteeing well-established operations under a given intrinsic $Q$. In this configuration, we can estimate the group index detuning from $10^3$ to $10^4$ with the switching speed from a few MHz to GHz, according to our proposal.

In conclusion, we showed a CRIT system tailored for programmable slow-light dynamics. By devising a building block formulated under the spinor representation of bright and dark modes, our design extends the framework of traditional EIT through generalized unitary operations. This approach enables full reconfigurability of the bandwidth, line-shape asymmetry, and slow-light functionalities. Both the dispersion analysis and time-domain dynamics demonstrate that this platform holds significant promise as a versatile building block for delay lines and frequency converters, which are essential for optical interconnects. The connection between the Fano-asymmetric EIT and bound states in the continuum (BIC)[56] motivates the design for reconfigurable resonances in integrated photonic platforms.

## Methods

**Transmission spectra.** The transmission spectrum of the CRIT lattice is calculated using the temporal coupled mode theory. Starting from the frequency-domain representation of the coupled-mode equations (Supplementary Note S1), the steady-state field amplitude in the bright-mode resonator, $\psi_B$, is

$$\psi_{\mathrm{B}}=\frac{i\left(\omega-\omega_{\mathrm{D}}\right)\sqrt{\frac{1}{\tau_{\mathrm{B}}{}^{\mathrm{W}}}}}{i\left(\omega-\omega_{\mathrm{D}}\right)\left[i\left(\omega-\omega_{\mathrm{B}}\right)+\frac{1}{2\tau_{\mathrm{B}}{}^{\mathrm{W}}}+\frac{1}{2\tau_{\mathrm{B}}{}^{\mathrm{I}}}\right]+\frac{1}{2}\frac{1}{\tau_{\mathrm{D}}\tau_{\mathrm{B}}}\left(1+\cos\left(\xi^{\mathrm{U}}-\xi^{\mathrm{L}}\right)\right)}\varphi_{\mathrm{I}}. \tag{4}$$

Substituting Eq. (4) into the input-output relation (Supplementary Note S1) yields

$$T=\left|\frac{\varphi_{\mathrm{O}}}{\varphi_{\mathrm{I}}}\right|^{2}=1-\frac{4\frac{\tau_{\mathrm{B}}{}^{\mathrm{W}}}{\tau_{\mathrm{B}}{}^{\mathrm{I}}}}{\left(1+\frac{\tau_{\mathrm{B}}{}^{\mathrm{W}}}{\tau_{\mathrm{B}}{}^{\mathrm{I}}}\right)^{2}+4\left[\left(\omega-\omega_{\mathrm{B}}\right)-\frac{1}{2}\frac{1}{\tau_{\mathrm{D}}\tau_{\mathrm{B}}}\frac{\left(1+\cos\left(\xi^{\mathrm{U}}-\xi^{\mathrm{L}}\right)\right)}{\left(\omega-\omega_{\mathrm{D}}\right)}\right]^{2}\left(\tau_{\mathrm{B}}{}^{\mathrm{W}}\right)^{2}}. \tag{5}$$

The calculated spectra are presented in Fig. 2.

**Spectral bandwidth.** The FWHM Γ shown in Fig. 2 characterizes the spectral linewidth of the transparency window. We calculate the FWHM under the condition of $\omega_B = \omega_D = \omega_0$, achieving

$$\Gamma=-C+\sqrt{C^{2}+4X}, \tag{6}$$

where $X = [1+\cos(\xi^U-\xi^L)]/(2\tau_D\tau_B)$ and $C = 8/(\tau_B{}^I\tau_B{}^W)(1/\tau_B{}^W+1/\tau_B{}^I)$.

**Spectral asymmetry.** The factor $F$ provides a measure of the spectral asymmetry in the transmission profile, which is defined as[11]:

$$F=\frac{\Delta\omega_{\text{high}}}{\Delta\omega_{\text{low}}}, \tag{7}$$

where $\Delta\omega_{\text{high}}$ and $\Delta\omega_{\text{low}}$ denote the frequency ranges from the central peak to the higher and lower half-maximum points, respectively.

**Numerical Simulation.** The simulation is performed using the 4th-order Runge–Kutta method with a temporal step size of $\Delta t = T_0/1024$. We consider a 1D lattice of 60-unit cells, where the inter-cell connection is characterized by a propagation delay of $1.25T_0$. The system is excited at the first unit cell by the input pulse field, $\varphi_I(t) = \exp(i\omega_0 t)\exp[-(t-t_c)^2/(2\sigma^2)]$, with a temporal width $\sigma = 100T_0$ and pulse centre $t_c = 4\sigma$. $\varphi_O(\omega)$ is obtained with the Fourier transform of $\varphi_O(t)$.

## Data availability

The data that support the plots and other findings of this study are available from the corresponding author upon request.

## Code availability

All code developed in this work will be made available upon request.

## Acknowledgements

We acknowledge financial support from the National Research Foundation of Korea (NRF) through the Basic Research Laboratory (No. RS-2024-00397664), Innovation Research Center (No. RS-2024-00413957), Young Researcher Program (No. RS-2025-00552989), and Midcareer Researcher Program (No. RS-2023-00274348), all funded by the Korean government. This work was supported by Creative-Pioneering Researchers Program and the BK21 FOUR program of the Education and Research Program for Future ICT Pioneers in 2025, through Seoul National University. We also acknowledge an administrative support from SOFT foundry institute.

## Author contributions

S.Y., X.P., and N.P. conceived the idea of bridging programmable photonics with slow light and supervised the project. S.P. conducted all the theoretical study and numerical analyses. B.C. contributed to the discussion on the practical implementation of the proposed design. All authors contributed to discussions of the results and to the writing of the manuscript.

## Competing interests

The authors have no conflicts of interest to declare.

# Supplementary Information for "Fully programmable slow light based on a spinor representation of generalized coupled-resonator-induced transparency"

Seungkyun Park[1,2], Beomjoon Chae[2], Hyungchul Park[2], Sunkyu Yu[2§], Xianji Piao[3†], and Namkyoo Park[1*]

[1]Photonic Systems Laboratory, Department of Electrical and Computer Engineering, Seoul National University, Seoul 08826, Korea

[2]Intelligent Wave Systems Laboratory, Department of Electrical and Computer Engineering, Seoul National University, Seoul 08826, Korea

[3]Wave Engineering Laboratory, School of Electrical and Computer Engineering, University of Seoul, Seoul 02504, Korea

E-mail address for correspondence: [§]sunkyu.yu@snu.ac.kr, [†]piao@uos.ac.kr, [*]nkpark@snu.ac.kr

**Supplementary Note S1. Tight-binding Hamiltonian of a CRIT unit cell**

In this note, we derive the tight-binding Hamiltonian of a CRIT unit cell. The temporal coupled mode equation for the structure described in Supplementary Fig. S1 becomes

$$\frac{d}{dt}\begin{bmatrix}\psi_{\mathrm{D}}\\ \psi_{\mathrm{B}}\end{bmatrix}=\begin{bmatrix} i\omega_{\mathrm{D}}-\frac{1}{2\tau_{\mathrm{D}}{}^{\mathrm{U}}}-\frac{1}{2\tau_{\mathrm{D}}{}^{\mathrm{L}}} & 0\\ 0 & i\omega_{\mathrm{B}}-\frac{1}{2\tau_{\mathrm{B}}{}^{\mathrm{U}}}-\frac{1}{2\tau_{\mathrm{B}}{}^{\mathrm{L}}}-\frac{1}{2\tau_{\mathrm{B}}{}^{\mathrm{W}}}-\frac{1}{2\tau_{\mathrm{B}}{}^{\mathrm{I}}}\end{bmatrix}\begin{bmatrix}\psi_{\mathrm{D}}\\ \psi_{\mathrm{B}}\end{bmatrix}$$
$$+\begin{bmatrix}\sqrt{\frac{1}{\tau_{\mathrm{D}}{}^{\mathrm{U}}}} & 0\\ 0 & \sqrt{\frac{1}{\tau_{\mathrm{B}}{}^{\mathrm{U}}}}\end{bmatrix}\begin{bmatrix}\mu_{\mathrm{DI}}{}^{\mathrm{U}}\\ \mu_{\mathrm{BI}}{}^{\mathrm{U}}\end{bmatrix}+\begin{bmatrix}\sqrt{\frac{1}{\tau_{\mathrm{D}}{}^{\mathrm{L}}}} & 0\\ 0 & \sqrt{\frac{1}{\tau_{\mathrm{B}}{}^{\mathrm{L}}}}\end{bmatrix}\begin{bmatrix}\mu_{\mathrm{DI}}{}^{\mathrm{L}}\\ \mu_{\mathrm{BI}}{}^{\mathrm{L}}\end{bmatrix}+\begin{bmatrix}0\\ \sqrt{\frac{1}{\tau_{\mathrm{B}}{}^{\mathrm{W}}}}\varphi_{\mathrm{I}}\end{bmatrix}, \tag{S1}$$

$$\begin{bmatrix}\mu_{\mathrm{DO}}{}^{\mathrm{U}}\\ \mu_{\mathrm{BO}}{}^{\mathrm{U}}\end{bmatrix}=\begin{bmatrix}\mu_{\mathrm{DI}}{}^{\mathrm{U}}\\ \mu_{\mathrm{BI}}{}^{\mathrm{U}}\end{bmatrix}-\begin{bmatrix}\sqrt{\frac{1}{\tau_{\mathrm{D}}{}^{\mathrm{U}}}} & 0\\ 0 & \sqrt{\frac{1}{\tau_{\mathrm{B}}{}^{\mathrm{U}}}}\end{bmatrix}\begin{bmatrix}\psi_{\mathrm{D}}\\ \psi_{\mathrm{B}}\end{bmatrix}, \tag{S2}$$

$$\begin{bmatrix}\mu_{\mathrm{DO}}{}^{\mathrm{L}}\\ \mu_{\mathrm{BO}}{}^{\mathrm{L}}\end{bmatrix}=\begin{bmatrix}\mu_{\mathrm{DI}}{}^{\mathrm{L}}\\ \mu_{\mathrm{BI}}{}^{\mathrm{L}}\end{bmatrix}-\begin{bmatrix}\sqrt{\frac{1}{\tau_{\mathrm{D}}{}^{\mathrm{L}}}} & 0\\ 0 & \sqrt{\frac{1}{\tau_{\mathrm{B}}{}^{\mathrm{L}}}}\end{bmatrix}\begin{bmatrix}\psi_{\mathrm{D}}\\ \psi_{\mathrm{B}}\end{bmatrix}, \tag{S3}$$

$$\begin{bmatrix}\mu_{\mathrm{DI}}{}^{\mathrm{U}}\\ \mu_{\mathrm{BI}}{}^{\mathrm{U}}\end{bmatrix}=\begin{bmatrix}0 & e^{-i\Phi_{\mathrm{DB}}{}^{\mathrm{U}}}\\ e^{-i\Phi_{\mathrm{BD}}{}^{\mathrm{U}}} & 0\end{bmatrix}\begin{bmatrix}\mu_{\mathrm{DO}}{}^{\mathrm{U}}\\ \mu_{\mathrm{BO}}{}^{\mathrm{U}}\end{bmatrix}, \tag{S4}$$

$$\begin{bmatrix}\mu_{\mathrm{DI}}{}^{\mathrm{L}}\\ \mu_{\mathrm{BI}}{}^{\mathrm{L}}\end{bmatrix}=\begin{bmatrix}0 & e^{-i\Phi_{\mathrm{DB}}{}^{\mathrm{L}}}\\ e^{-i\Phi_{\mathrm{BD}}{}^{\mathrm{L}}} & 0\end{bmatrix}\begin{bmatrix}\mu_{\mathrm{DO}}{}^{\mathrm{L}}\\ \mu_{\mathrm{BO}}{}^{\mathrm{L}}\end{bmatrix}, \tag{S5}$$

$$\varphi_{\mathrm{O}}=\varphi_{\mathrm{I}}-\sqrt{\frac{1}{\tau_{\mathrm{B}}{}^{\mathrm{W}}}}\psi_{\mathrm{B}}, \tag{S6}$$

where $\mu_{\mathrm{DI}}{}^{\mathrm{U,L}}$, $\mu_{\mathrm{DO}}{}^{\mathrm{U,L}}$, $\mu_{\mathrm{BI}}{}^{\mathrm{U,L}}$, and $\mu_{\mathrm{BO}}{}^{\mathrm{U,L}}$ are the field amplitude at each position of the upper ('U') or lower ('L') waveguide loops, and $\Phi_{BD}{}^{U,L}$ and $\Phi_{DB}{}^{U,L}$ are the phase shifts between the bright- and dark-mode resonators along the upper ('U') or lower ('L') waveguide loops. By applying Eqs. (S2)–(S5), the field amplitudes within the upper and lower waveguide loops are expressed as follows:

$$\begin{bmatrix} \mu_{DI}{}^{U} \\ \mu_{BI}{}^{U} \end{bmatrix} = \frac{1}{2}\begin{bmatrix} 1 & e^{i\Phi_{BD}{}^{U}} \\ e^{i\Phi_{DB}{}^{U}} & 1 \end{bmatrix}\begin{bmatrix} \sqrt{\frac{1}{\tau_D{}^{U}}} & 0 \\ 0 & \sqrt{\frac{1}{\tau_B{}^{U}}} \end{bmatrix}\begin{bmatrix} \psi_D \\ \psi_B \end{bmatrix},$$

$$\begin{bmatrix} \mu_{DI}{}^{L} \\ \mu_{BI}{}^{L} \end{bmatrix} = \frac{1}{2}\begin{bmatrix} 1 & e^{i\Phi_{BD}{}^{L}} \\ e^{i\Phi_{DB}{}^{L}} & 1 \end{bmatrix}\begin{bmatrix} \sqrt{\frac{1}{\tau_D{}^{L}}} & 0 \\ 0 & \sqrt{\frac{1}{\tau_B{}^{L}}} \end{bmatrix}\begin{bmatrix} \psi_D \\ \psi_B \end{bmatrix}. \tag{S7}$$

Substituting Eq.(S7) into Eq. (S1) yields the following equation.

$$\begin{aligned} \frac{d}{dt}\begin{bmatrix} \psi_D \\ \psi_B \end{bmatrix} = & \begin{bmatrix} i\omega_D - \frac{1}{2\tau_D{}^{U}} - \frac{1}{2\tau_D{}^{L}} & 0 \\ 0 & i\omega_B - \frac{1}{2\tau_B{}^{U}} - \frac{1}{2\tau_B{}^{L}} - \frac{1}{2\tau_B{}^{W}} - \frac{1}{2\tau_B{}^{I}} \end{bmatrix}\begin{bmatrix} \psi_D \\ \psi_B \end{bmatrix} \\ & + \frac{1}{2}\begin{bmatrix} \sqrt{\frac{1}{\tau_D{}^{U}}} & 0 \\ 0 & \sqrt{\frac{1}{\tau_B{}^{U}}} \end{bmatrix}\begin{bmatrix} 1 & e^{i\Phi_{BD}{}^{U}} \\ e^{i\Phi_{DB}{}^{U}} & 1 \end{bmatrix}\begin{bmatrix} \sqrt{\frac{1}{\tau_D{}^{U}}} & 0 \\ 0 & \sqrt{\frac{1}{\tau_B{}^{U}}} \end{bmatrix}\begin{bmatrix} \psi_D \\ \psi_B \end{bmatrix} \\ & + \frac{1}{2}\begin{bmatrix} \sqrt{\frac{1}{\tau_D{}^{L}}} & 0 \\ 0 & \sqrt{\frac{1}{\tau_B{}^{L}}} \end{bmatrix}\begin{bmatrix} 1 & e^{i\Phi_{BD}{}^{L}} \\ e^{i\Phi_{DB}{}^{L}} & 1 \end{bmatrix}\begin{bmatrix} \sqrt{\frac{1}{\tau_D{}^{L}}} & 0 \\ 0 & \sqrt{\frac{1}{\tau_B{}^{L}}} \end{bmatrix}\begin{bmatrix} \psi_D \\ \psi_B \end{bmatrix} + \begin{bmatrix} 0 \\ \sqrt{\frac{1}{\tau_B{}^{W}}}\varphi_I \end{bmatrix}. \end{aligned} \tag{S8}$$

To achieve the zero-field condition inside the loop couplers, we set $\Phi_{BD}{}^{U,L} = 2q_{BD}{}^{U,L}\pi + \pi/2 + \xi^{U,L}$ and $\Phi_{DB}{}^{U,L} = 2q_{DB}{}^{U,L}\pi + \pi/2 - \xi^{U,L}$ to satisfy $\Phi_{BD}{}^{U,L} + \Phi_{DB}{}^{U,L} = (2q+1)\pi$ (where $q$ = 0, 1, 2,…), where $q_{BD}^{U,L}$ and $q_{DB}^{U,L}$ are nonnegative integers. Assuming the symmetric coupling condition, $\tau_D^U = \tau_D^L = \tau_D$ and $\tau_B^U = \tau_B^L = \tau_B$, Eq. (S8) becomes:

$$\frac{d}{dt}\begin{bmatrix}\psi_D \\ \psi_B\end{bmatrix} = \begin{bmatrix} i\omega_D & \frac{i}{2}\sqrt{\frac{1}{\tau_D\tau_B}}\left(e^{-i\xi^U}+e^{-i\xi^L}\right) \\ \frac{i}{2}\sqrt{\frac{1}{\tau_D\tau_B}}\left(e^{i\xi^U}+e^{i\xi^L}\right) & i\omega_B - \frac{1}{2\tau_B^W} - \frac{1}{2\tau_B^I}\end{bmatrix}\begin{bmatrix}\psi_D \\ \psi_B\end{bmatrix} + \begin{bmatrix} 0 \\ \sqrt{\frac{1}{\tau_B^W}}\varphi_I\end{bmatrix}. \quad \text{(S9)}$$

By applying the spinor representation, $\Psi = [\psi_D, \psi_B]^T$, Eq. (S9) can be rewritten as:

$$\begin{aligned} i\frac{d}{dt}|\Psi\rangle = &-\left[\left(\omega_0 + \frac{i}{4\tau_B^I}\right)I + \left(\Delta\omega + \frac{i}{4\tau_B^I}\right)\sigma_z\right]|\Psi\rangle \\ &-\frac{1}{2}\sqrt{\frac{1}{\tau_D\tau_B}}\left[\left(\cos\xi^U + \cos\xi^L\right)\sigma_x + \left(\sin\xi^U + \sin\xi^L\right)\sigma_y\right]|\Psi\rangle, \end{aligned} \quad \text{(S10)}$$

which leads to Eqs. (2) and (3) in the main text.

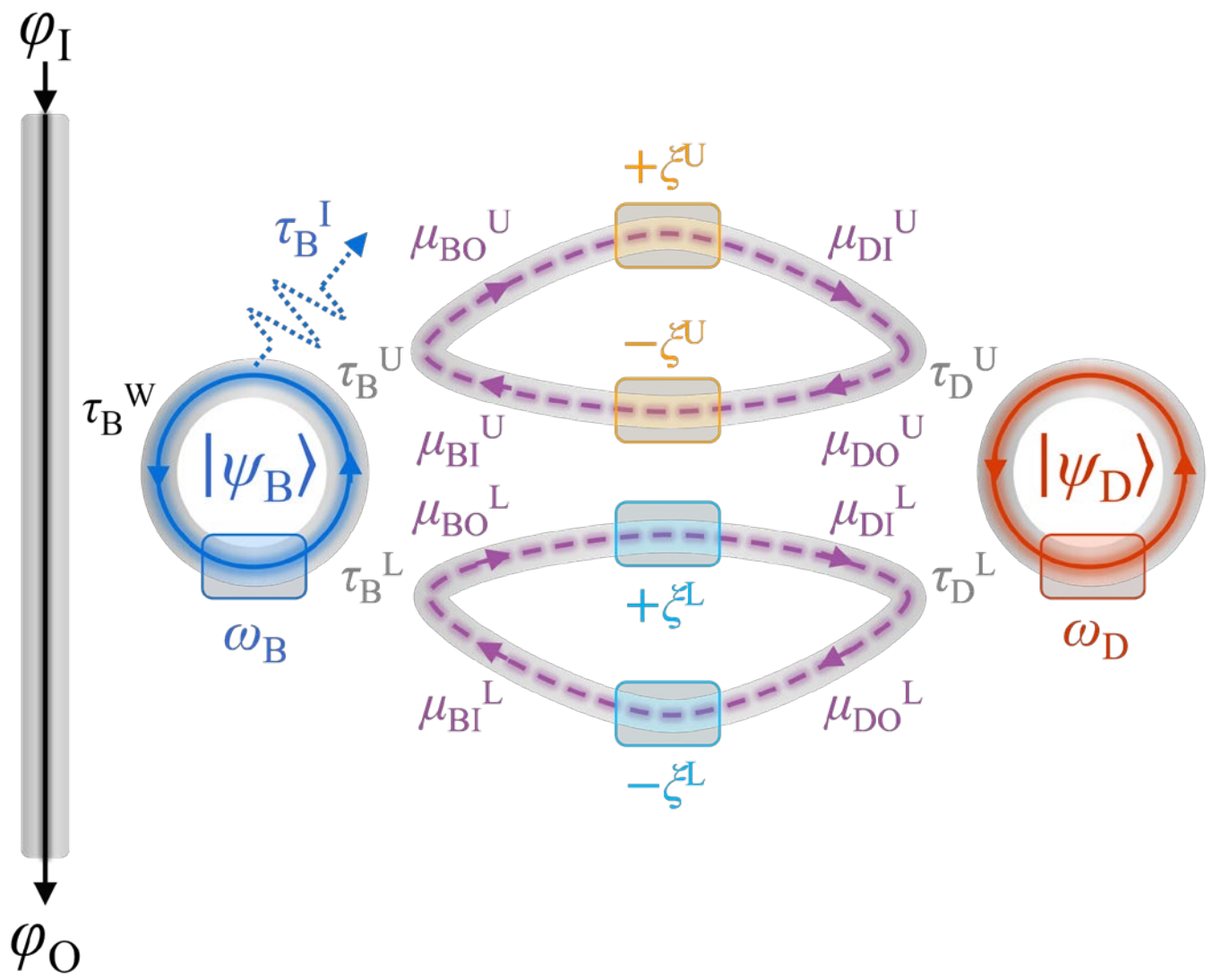

**Supplementary Figure S1. Programmable unit cell for generalized spinor-based CRIT.**

**Supplementary Note S2. Dispersion relation and group velocity**

To characterize the propagation dynamics of the CRIT lattice, we derive the dispersion relation and group velocity based on temporal coupled mode theory under the Bloch theorem. For incident and outgoing waveguide fields, $\varphi_{\mathrm{I},m}$ and $\varphi_{\mathrm{O},m}$, at the $m$-th unit cell, the propagation along the waveguide yields the phase evolution of $\varphi_{\mathrm{I},m} = \varphi_{\mathrm{O},m-1}e^{-i\beta\Lambda}$, where $\beta$ is the propagation constant. The Bloch theorem, $\varphi_{\mathrm{I},m} = \varphi_{\mathrm{I},m-1}\exp(-ik_x\Lambda)$, for the Bloch wavevector $k_x$ leads to the following relation:

$$i(\omega-\omega_{\mathrm{D}})\left[i(\omega-\omega_{\mathrm{B}})+\frac{1}{\tau_{\mathrm{B}}{}^{\mathrm{W}}}\left(1+\frac{2e^{ik'\Lambda}}{1-e^{ik'\Lambda}}\right)\right]+\frac{1}{2}\frac{1}{\tau_{\mathrm{D}}\tau_{\mathrm{B}}}\left(1+\cos\left(\xi^{\mathrm{U}}-\xi^{\mathrm{L}}\right)\right)=0. \quad \text{(S11)}$$

where $k' = k_x - \beta$ represents the relative Bloch wavevector. Solving this equation for $k_x$ yields the following dispersion relation:

$$k_x = \beta - \frac{i}{\Lambda}\log\left(\frac{A+1}{A-1}\right), \quad \text{(S12)}$$

where the auxiliary function $A$ is defined as

$$A = \left[i(\omega-\omega_{\mathrm{B}}) - \frac{i}{(\omega-\omega_{\mathrm{D}})}\frac{1}{2}\frac{1}{\tau_{\mathrm{D}}\tau_{\mathrm{B}}}\left(1+\cos\left(\xi^{\mathrm{U}}-\xi^{\mathrm{L}}\right)\right)\right]\tau_{\mathrm{B}}{}^{\mathrm{W}}. \quad \text{(S13)}$$

The group velocity $v_{\mathrm{g}}$ is obtained from Eq. (S13) as follows:

$$v_g = \left(\frac{\partial k_x}{\partial \omega}\right)^{-1} = \frac{1}{\frac{n_{\mathrm{eff}}}{c} - \frac{i}{\Lambda}\frac{2}{1-A^2}\frac{dA}{d\omega}}. \quad \text{(S14)}$$

This expression reveals that the group velocity depends on $\xi$, enabling reconfigurable group velocity using tunable phase shifters.

## Supplementary Note S3. Effect of material loss

In the main text, we assume the lossless condition for the bright-mode resonators. To consider a more practical scenario, we calculate the complex-valued dispersion band in Fig. S2 for the finite intrinsic $Q$-factor of bright-mode resonators with $\tau_B^I = 20 \times 2\pi/\omega_0$. The result shows that the decay of optical modes is enhanced at the slow-light bands as expected from their strong light-matter interactions.

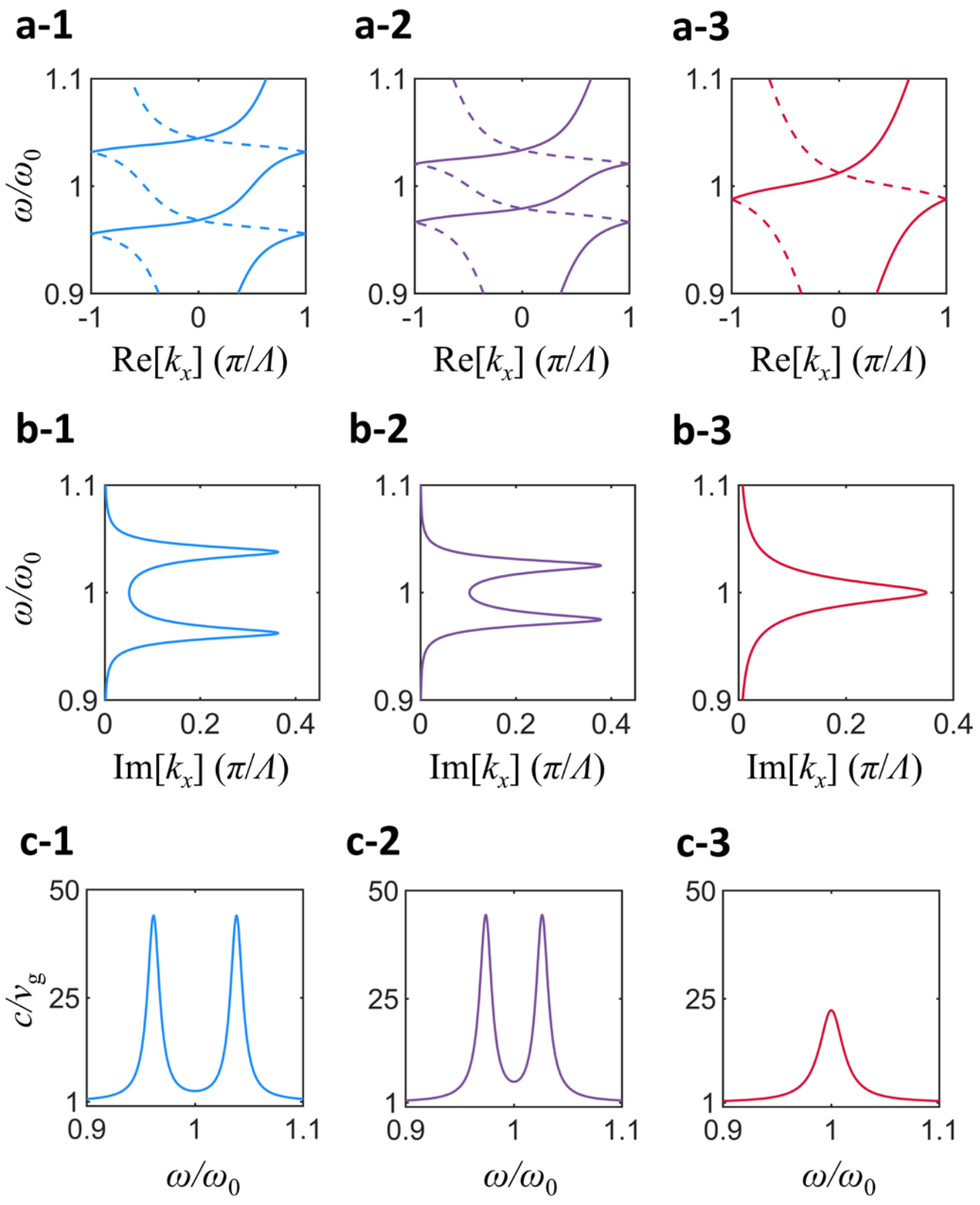

**Supplementary Figure S2. Complex dispersion relations with finite resonator lifetimes. a,b** Complex-valued dispersion relations for real (**a**) and imaginary (**b**) parts. (**c**) The corresponding group indices under $x$-axis rotations. $\tau_B^I = 20 \times 2\pi/\omega_0$. All the other parameters are the same as those in Fig. 3 in the main text.

**Supplementary Note S4. Bandwidth dependency of modulation speed**

Although we assume an ideal stepwise modulation of the system parameters in Fig. 4 of the main text, the finite characteristic time required to dynamically reach the CRIT steady state prevents an instantaneous response of the group velocity to the applied modulation. Notably, a narrower spectral bandwidth centred at $\omega_0$ leads to an extended characteristic time—consistent with the largest group velocity at $\omega_0$ shown in Fig. 3c-3 in the main text—resulting in a slower temporal response (Supplementary Fig. S3): $\Delta(L/\Lambda) = 22$ (illustrated by the distance between dashed lines), representing a twofold increase compared to $\Delta(L/\Lambda) = 11$ in Fig. 4a of the main text.

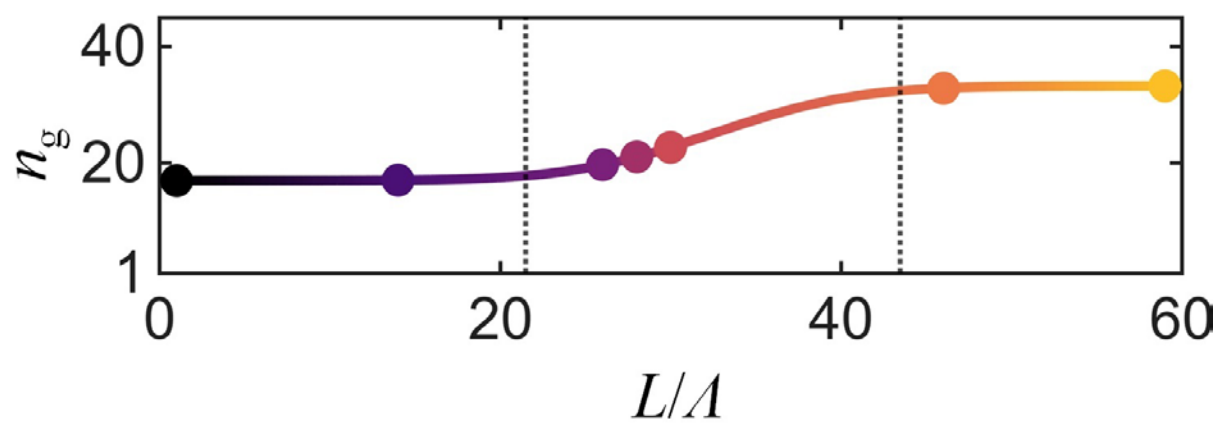

**Supplementary Figure S3. Bandwidth dependency of group index dynamics.** Group index $n_g$ evaluated at each unit cell along the 1D programmable CRIT lattice for a pulse bandwidth of $\Delta\omega_0/\omega_0 = 1.325\times10^{-3}$, which is half the value used in Fig. 4a of the main text. The gauge field $\xi$ is switched from $\pi/2$ to 0 at $t = 1500T_0$. All the other parameters are the same as those in Fig. 4a of the main text.